\documentstyle[prb,aps,multicol,epsfig]{revtex}
\tighten
\newcommand{\beq}{\begin{equation}}
\newcommand{\eeq}{\end{equation}}
\newcommand{\bea}{\begin{eqnarray}}
\newcommand{\eea}{\end{eqnarray}}

\begin{document}

\draft

\title{A model of inversion of DNA charge by a positive 
polymer: fractionization of the polymer charge}

\author{T. T. Nguyen and B. I. Shklovskii}

\address{Theoretical Physics Institute, University of Minnesota, 116
  Church St. Southeast, Minneapolis, Minnesota 55455}

\maketitle

\vspace{.2cm}
\centerline{\today}

\begin{abstract}
Charge inversion of a DNA double helix by an oppositely 
charged flexible polyelectrolyte (PE) is considered.
We assume that,
in the neutral state of the DNA-PE complex,
each of the DNA charges is
locally compensated by a PE charge. 
When an additional PE molecule is adsorbed
by DNA, its charge gets fractionized into monomer charges of 
defects (tails and arches) on the background of the perfectly 
neutralized DNA.
These charges
spread all over the DNA eliminating the self-energy of PE. 
This fractionization mechanism 
leads to a substantial inversion of the DNA charge, 
a phenomenon which is widely used for gene delivery.  
\end{abstract}

\begin{multicols} {2}

Inversion of the negative charge 
of a DNA double helix by its complexation with
a positive polyelectrolyte (PE) is 
used for the gene delivery. 
The positive charge of DNA-PE
complex facilitates DNA contact with a typically
negative cell membrane making penetration into the cell 
hundreds times more likely\cite{Kabanovs}. Charge inversion of DNA-PE
complexes was confirmed recently by 
electrophoresis\cite{Kabanov}.
If at a given concentration of long DNA helices one increases
concentration of shorter PE molecules, at some critical point the  
electrophoretic mobility of a DNA-PE complex 
changes sign from negative to positive.
Intuitively, one can think that when a 
PE completely neutralizes a DNA double helix
new molecules of PE do not attach to DNA.
Indeed, the Poisson-Boltzmann approximation
for description of screening of a DNA 
by any counterions including 
PE does not lead to charge inversion. 
Counterintuitive 
phenomenon of charge inversion of a macroion by 
oppositely charged PE
has attracted significant 
attention\cite{Linse,Pincus,Bruinsma,Joanny,Joanny1,Joanny2,Nguyen,Nguyen1,Nguyen2,Nguyen3,Rubinstein,Shklovskii,RMP}. 
It can be explained 
if one takes into account that the surface potential
of already neutralized DNA is locally affected by 
a new approaching PE molecule, or in other words, taking into
account correlations between PE molecules\cite{Shklovskii,RMP}. 
Due to repulsive interaction between PE molecules
a new PE molecule pushes aside 
already adsorbed on DNA surface 
molecules and creates on the surface 
an oppositely charged
image of itself. The image attract the new PE 
molecule leading to charge inversion.
This phenomenon 
is similar to attraction of a charge 
to a neutral metal.
For quantitative consideration 
charges of DNA are often assumed to be smeared and 
to form uniformly charged 
cylinder\cite{Linse,Pincus,Bruinsma,Joanny,Joanny1,Joanny2,Nguyen,Nguyen1,Nguyen2,Nguyen3,Rubinstein,Shklovskii,RMP}. 
This approach ignores interference between 
chemical structure of DNA surface and of PE
and clearly is not fully satisfactory.
In this paper, we consider effects of discreteness and 
configuration of $-e$ charges 
of the DNA double helix. In this case, we
suggest an explanation of charge inversion
based on ``fractionization" of charge of PE molecules.
It turns out to be even simpler 
and more visual than for smeared charges
of DNA.

\begin{figure}
\epsfxsize=8cm \centerline{\epsfbox{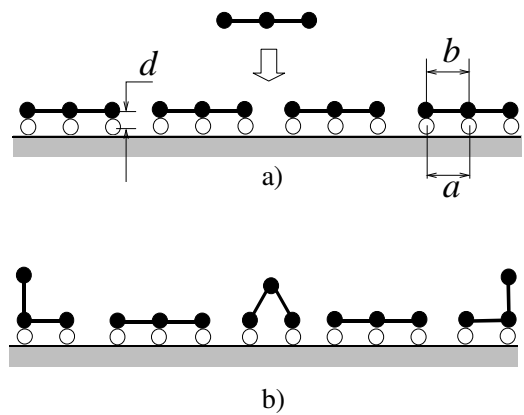}}
\caption{The origin of charge fractionization in PE 
adsorption. a) One of spirals of negative charges of DNA 
(empty circles) is completely 
neutralized by positive PE molecules with $Z=3$
(their monomers are shown by solid circles). A new 
PE molecule is approaching DNA. b) The new PE molecule is "digested" 
by DNA. Its charge is split in $+e$ charges of $Z$ defects.
They are tails and an arch (center). }
\label{fig:model}
\end{figure}

Negative elementary
charges of DNA phosphates are situated along the two 
spirals at the exterior of both helices.
When unfolded, each spiral is an one-dimensional 
lattice of such charges, with the lattice constant $a$=6.7\AA.
Let us consider a toy model of a PE as a freely jointed chain of
$Z$ small $+e$ monomers. The elastic energy cost for bending
the PE is neglected in this model, so that one can concentrate 
on the electrostatic aspect of the problem. 
To maximize the role of
discreteness of DNA charge we assume that 
the PE bond length $b$ is exactly equal to
the distance $a$ between negative charges of a spiral.
(The case when these lengths are different
is discussed in the end of the paper).
We assume that minimal distance, $d$, between 
a PE charge and a charge of DNA is much smaller than $a$.
Then  PE molecules can attach to a DNA charge spiral
in such a way that 
every charge of a spiral is locally compensated by a PE charge and, 
therefore, DNA is completely neutralized. 
The case of $Z =3$ is shown in Fig.  \ref{fig:model}a.
The neutralization is so perfect that
it is difficult to imagine how another PE molecule 
can be attached to DNA.
%}
In this paper, we would like to discuss the fractionization
mechanism which brings an additional PE to the
neutralized DNA and leads to charge inversion.
Fig.  \ref{fig:model} shows how this mechanism works
for the case of $Z =3$. 
When a new PE comes to the DNA double 
helix which is already neutralized
by PE, it creates a place for itself or, in other words,
the oppositely charged image
in the following way. Let us choose 
$Z$ already-absorbed PE molecules, which 
are situated far from each other. 
In each of them we detach one PE monomer from DNA 
surface. This leads to 
formation of positive defects 
(tails and arches) and negative 
vacancies on DNA.
To create an image for a new 
PE let us shift adsorbed PE molecules along DNA 
in such a way that all $Z$ vacancies join together and 
form a large vacancy of a length $Z$. A new 
PE molecule is accommodated in this vacancy.
As a result of consumption of this molecule 
$Z$ defect $+e$ charges appear on the top 
of completely neutralized spiral 
(see Fig.  \ref{fig:model}b).

This effectively looks as
cutting of the new PE molecule into 
$Z$ individual monomers and spreading
them out along the spiral. In other 
words, charge inversion of DNA happens by
fractionization of the PE molecule charge.
Of course, none of the chemical bonds 
is really cut, and this phenomenon
is solely due 
to the correlated distribution
of PE molecules, which avoid each other 
at the DNA spiral. In this sense,
fractionization we are talking about is 
similar to what happens in fractional quantum Hall
effect\cite{Tao} or in the polyacetilene\cite{Braz},
where many-electron correlations result in fractionization of 
the electron charge.

Fractionization is driven by elimination 
of the self-energy of free PE in solution.
By the self-energy we mean the energy of
repulsive interactions of $Z$
positive charges of the PE molecule
in extended conformation which it has in the solution.
In the fractionized state, charges of monomers 
are far from each other and practically do not
interact, so that the PE
self-energy is eliminated and, therefore, gained. 

Let us now calculate 
the net inverted charge using this 
fractionization mechanism. 
We denote the linear charge density
of the inverted (positive) 
net charge of the double helix DNA by $\eta^*$.
The chemical potential of the PE absorbed
at the spiral is
\beq
\mu_s=Zk_BT\ln(\eta^*/\eta_0)+Ze \psi(0)~.
\label{eq:muabsorbed}
\eeq
The first term in the right hand side
of Eq. (\ref{eq:muabsorbed}) is
the chemical potential of
the one-dimensional gas of defects
($-\eta_0 \simeq 0.6e/$\AA\ is the bare charge density of DNA).
We used expression for the chemical potential of an
ideal gas because the Coulomb
interaction between defects at the a distance of a
few $a$ is much smaller than $k_BT$ ($a \simeq l_B$, 
where $l_B = e^2/Dk_BT \simeq 7$\AA\ is the Bjerrum length.) 
The second term in the right hand 
side of Eq. (\ref{eq:muabsorbed}) 
is the repulsion energy of the new PE
from the inverted charge of the DNA. 
In this term, $\psi(0)$ is
the averaged surface potential of the DNA helix. 
We assume in this paper that the net charge 
of DNA is screened by a monovalent salt
at the screening length $r_s$, which is much larger than $a$.
Then $\psi(0)$ can be calculated as
the surface potential
of a cylinder with radius of DNA helix $R$ 
and linear density of charge $\eta^*$
\beq
\psi(0)\simeq \frac{2\eta^*}{D}\ln\frac{r_s+R}{R}~.
\eeq
To find $\eta^*$ in the equilibrium state, 
one has to equate the chemical potential of the absorbed
PE molecules with that of the free PE in the solution. 
The later one can be
calculated as following. Due to the
repulsive Coulomb interaction between monomers,
a free PE in the solution has an extended shape to minimize its energy.
Therefore, the chemical potential
of a free PE in solution can be written as the self-energy of 
a rigid rod with the length $Na$ and the linear charge density $e/a$
\beq
\mu_0= (Ze^2/Da)\ln({\cal L}/a)~,
\label{eq:mufree}
\eeq
where ${\cal L}=\min(r_s,Za)$ and 
$D$ is the dielectric constant of water.
We have assumed that the concentration 
of PE in the solution is large enough
so that its translational entropy can 
be neglected. In this sense,
we are calculating the maximum possible charge inversion. 
If the PE molecule is long $(Z\gg 1)$ this limit is reached at
a concentration of PE which is exponentially small
$(\sim \exp(-Z))$.

Equating the chemical potentials of
Eqs. (\ref{eq:mufree}) and  (\ref{eq:muabsorbed}),
one has
\beq
\psi(0) =(e/Da)\ln({\cal L}/a)+ (k_BT/e)\ln(\eta_0/\eta^*)~.
\label{eq:charging}
\eeq
One can interpret the right hand 
side as a ``correlation" voltage 
(provided by the total free energy gain
in fractionization of PE charge) that
(over-)charges the DNA to the potential $\psi(0)$.

To the first approximation, one can neglect
the entropic term in the
right hand side of Eq. (\ref{eq:charging})
 and easily get
a solution for the net charge density
\beq
\eta^*\simeq\frac{e}{2a}\frac{\ln({\cal L}/a)}{\ln[(r_s+R)/R]}~.
\eeq
Now one can check that this solution
is consistent with the assumption
that the entropic term can be 
neglected by substituting it back into
Eq. (\ref{eq:charging}). 
Of course, $\eta^*$ is positive 
indicating that the bare DNA
charge is inverted. 
Knowing $\eta^*$ and using $|\eta_0| = 0.6e/$\AA$ \simeq 3.9 e/a$
the charge inversion
ratio can be calculated
\beq
\left|\frac{\eta^*}{\eta_0}\right| = 
0.13 \frac{\ln({\cal L}/a)}{\ln[(r_s+R)/R]} ~.
\eeq

For DNA $R = 10$\AA\ and $a = 6.7$\AA, so that at $r_s \geq 10$\AA 
the ratio of logarithms 
can be only slightly larger than unity. Thus, the
charge inversion ratio created by fractionization is limited by 
20\%.  Up to such point we indeed can neglect
Coulomb interactions between defects in the chemical potential
of the gas of defects (the first term in the right hand side
of Eq. (\ref{eq:muabsorbed}).

We emphasize that
it would be incorrect to conclude that 
fractionization discussed above is a strictly  
one-dimensional phenomenon,
similarly to the well 
known cases of one-dimensional 
density wave\cite{Tao} and 
polyacetilene\cite{Braz}. 
It is easy to see that our
fractionization  
mechanism applies equally well for
a two-dimensional surface with discrete 
charges which form a square
lattice with the same lattice constant as
the PE bond length $c$. Indeed,
one can view Fig. \ref{fig:model} as a 
cross-section of this lattice
and all previous arguments 
about the role of tails and arches 
can be carried over to this
case. There are, however, small modifications of the 
analytic formulae for charge inversion.
Defects with $+e$ charges 
form now a two-dimensional gas with
concentration
$\sigma^*/e$, where $\sigma^*$ is the net
positive surface charge density playing the
roles of $\eta^*$.
The chemical 
potential of this gas is $k_BT\ln(a^2\sigma^*/e)$.
The surface potential is now 
$\psi(0)=2\pi\sigma^* r_s/D$.
The balance of the chemical potential of PE molecules
adsorbed at the surface with that of a free PE 
in the solution now reads
\beq
2\pi\sigma^*r_s/D=(e/Da)\ln({\cal L}/a)+(k_BT/e)\ln(e/a^2\sigma^*)
\eeq
and the solution, for $a \simeq l_B$, within a numerical factor, is
\beq
\sigma^* \simeq (e/ar_s)/\ln(r_s/a)~.
\eeq
One can see that, for $a \simeq l_B$, in the free
energy gained by fractionization of 
the PE molecule charge, the entropy
contribution is comparable to the
self-energy, in contrary with
the one-dimensional case, where the entropic term can be neglected.
This is obviously due to a higher number of degrees of freedom 
which a 
two-dimensional surface provides to the gas of defects.
If $r_s \gg a$, the fractionization induced charge 
inversion ratio for the two-dimensional 
case is smaller than for DNA:
\beq
\left|\frac{\sigma^*}{e/a^2}\right|
=\frac{a}{r_s}\ln\frac{r_s}{a} ~.
\eeq
An important role of elimination of the self-energy 
for adsorption of a flexible 
PE on an oppositely uniformly charged surface was 
previously emphasized in Refs. \onlinecite{Pincus,Joanny}.

Until now we considered adsorption 
of linear charged molecules (PE)
both on one- and two-dimensional lattices of the
background charge.
It is interesting to note that 
the fractionization mechanism works 
for molecules of other shapes, too.
Let us, for example, consider dendrimers 
(star-like branching molecules 
with  a large number of monovalent positive charges 
on their periphery), which were
also shown to invert the charge of DNA
\cite{KabanovD}. Dendrimers
with charges $Z$=4, 8 can easily compensate
a compact group of nearest $Z$ charges of both DNA helices.
If a DNA double helix is totally covered and neutralized
by such dendrimers 
an additional dendrimer can still be adsorbed on 
DNA because one monomer of each of $Z$ already adsorbed 
dendrimers can be raised above the DNA surface.
As in the case of linear molecules, this leads to 
fractionization of the dendrimer charge
and to the gain of its self-energy.

Returning  to DNA-PE complexes we would like 
to remind the reader about
additional
mechanisms of charge inversion beside the fractionization
mechanism.
So far, to clearly demonstrate the
role of the fractionization mechanism in
charge inversion, we worked with the case when
distance between charges of PE, $b$, is equal 
to the distance between charges of an unfolded DNA spiral, $a$.
One can show that if $b$ varies in the vicinity of $a$,
the point $b = a$ is the local minimum 
of the charge inversion ratio.
Away from $b = a$ point, interaction
of a long PE molecule with
a spiral of DNA charges can be calculated
neglecting discreteness of PE and DNA
charges, i. e. assuming that the charge
is uniformly distributed
along both the PE molecule and the DNA spiral.

Let us first assume that $b < a$ so that the PE molecule has
larger linear charge density than the DNA spiral.
Then PE molecules repel each other and form on DNA
spiral a strongly correlated liquid where PE molecules alternate with
vacant places. This liquid reminds a Wigner crystal. 
A new PE molecule pushes aside previously adsorbed ones
or, in other words, attracts vacancies.
This provides another mechanism of creation of 
image of an approaching PE molecule in addition to the
defect formation mechanism.
Therefore, the negative chemical potential
of PE adsorbed on the spiral becomes larger 
in the absolute value
and charge inversion increases.
This is the same mechanism of 
Wigner-crystal-like correlations 
which was studied
in Ref. \onlinecite{Nguyen,Nguyen1,Shklovskii}.
In the opposite case, $b > a$, when PE 
has
a smaller linear charge density than a DNA spiral,
more than one layer of
PE is adsorbed on DNA to neutralize it.
Polarization
of the incomplete second layer by a new PE molecule
results again in an additional to defect formation
mechanism of attraction of this molecule to
a neutralized DNA\cite{Nguyen1}. This leads to
larger charge inversion than in $b = a$ case.

Changing flexibility of a PE we can separate
the role of fractionization. For example,
for the absolutely rigid PE molecules 
defects can not appear and fractionization
mechanism does not work. 
As a result at $b = a$, when the
layer of PE neutralizing DNA is completely incompressible
charge inversion vanishes 
(see a two-dimensional analog of this problem 
in Ref. \onlinecite{Nguyen3}).
In a flexible PE, 
the fractionization mechanism adds charge
inversion weakly dependent on $b$,
making the minimal value of the charge inversion ratio 
at $b = a$ finite.

Wigner-crystal-like correlations 
play an important role in the discussed above 
case of DNA charge inversion by dendrimers, too.
This happens when we deal with high generations of 
dendrimers which have a very large charge such as $32e$ or $64e$. 
Because of the three-dimensional architecture of 
their chemical bonds these molecules can 
not expand enough so that 
each charge of them reaches an opposite charge of DNA
and compensates it. In other words,
when projected to a DNA double helix, these high generation dendrimers
have much larger linear density
of charge than the double helix itself. Thus, 
large segments of the helix 
between adsorbed dendrimers remain negatively charged, 
and form a Wigner-Seitz cell around each dendrimer.
This is how with growing charge 
of dendrimers the fractionization mechanism gets replaced by the
mechanism of Wigner-crystal-like correlations.
Qualitative difference between low 
and high generations of dendrimers 
has been clearly demonstrated experimentally\cite{KabanovD}.

\begin{figure}
    \epsfxsize=8cm \centerline{\epsfbox{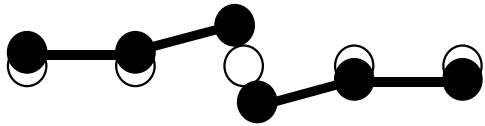}}
    \caption{A view from the top on a spiral of negative charges 
	of DNA (empty 
    circles) and two PE molecules. Two positive end monomers 
	share a negative charge of DNA in the perfect 
	opposition.}
    \label{fig:twoplus}
\end{figure}

Let us return to complexation of a DNA 
double helix with PE molecules with the matching
bond length, $b = a$, and discuss another
mechanism of charge inversion, 
which is related to the discreteness
of DNA charge and further increases 
the positive charge of DNA-PE complex. 
Let us consider a monomer tail of PE and explore 
whether some energy 
can be gained if the positive charge of this monomer 
moves down to the plane of DNA charges,
approaches already neutralized negative charge of the DNA and  
shares it with the end monomer of the neighboring PE molecule 
in a way shown in Fig.  \ref{fig:twoplus}.
If these two end monomers may sit on opposite sides of 
the negative charge of DNA, the 
additional energy $e^2/2d$ can be gained,
where $d$ is the distance of the 
closest approach of a PE monomer and a DNA charge\cite{Nguyen}.
At a sufficiently small $d$ this energy
can be even larger 
than the gain per tail from elimination of the self-energy.
In a DNA double helix, all the
negative charges indeed are on the ridge above 
neighboring neutral atoms. Two sufficiently small 
monomers may fit into 
the large and small groves on both sides of the ridge.
On the other hand if, because of sterical limitations,
they can 
not be in the perfect opposition the 
energy gain is smaller and can even vanish.
In this paper,
we have considered three mechanisms 
of charge inversion
of DNA by a PE:
fractionization of the PE charge, 
Wigner-crystal-like correlations and 
sharing of one DNA charge by two 
monomers of the PE. We showed that 
depending on properties of PE
they can work in different combinations.
In conclusion we emphasize that all these
mechanisms are due to the fact that a new 
PE molecule rearranges already
adsorbed PE in such away that its 
image or correlation hole strongly 
attracts this new PE molecule. 
None of these mechanisms can be described 
by the Poisson-Boltzmann
theory because this theory uses
the mean-field potential which 
does not depend on the position of the new PE molecule.
Thus all three effects are based on correlations 
between different PE molecules.
Fractionization of PE charge is the most
visual realization of these correlations. 

We are grateful to A. V. Kabanov for 
the question which initiated this work, 
and A.\ Yu.\ Grosberg and  
P.\ Pincus for useful discussions of results.
This work was supported by the NSF grant No. DMR-9985785.

\end{multicols}

\begin{references}

\bibitem{Kabanovs} A.\ V.\ Kabanov, V.\ A.\ Kabanov, 
Bioconjug. Chem. {\bf 6}, 7 (1995); 
Advanced Drug Delivery Reviews {\bf 30}, 49 (1998). 
\bibitem{Kabanov} V. A. Kabanov, A.\ A.\ Yaroslavov, S.\ A.\ Sukhisvili,
J. of Control Release, {\bf 39}, 173 (1996). 
\bibitem{Linse} T.\ Wallin, P.\ Linse, J. Phys. Chem. {\bf 100}, 
17873 (1996).
\bibitem{Pincus} E.\ M.\ Mateescu, C.\ Jeppersen and P.\ Pincus,
Europhys. Lett. {\bf 46}, 454 (1999).
\bibitem{Bruinsma} S.\ Y. Park, R.\ F.\ Bruinsma, and W.\ M.\ Gelbart,
Europhys. Lett. {\bf 46}, 493 (1999).
\bibitem{Joanny} J.\ F.\ Joanny, Europ. J. Phys. B {\bf 9} 117 (1999).
\bibitem{Joanny1} R. R. Netz, J.\ F.\ Joanny, Macromolecules, {\bf 32},
9013 (1999).
\bibitem{Joanny2} R. R. Netz, J.\ F.\ Joanny, Macromolecules, {\bf 32},
9026 (1999).
\bibitem{Nguyen} T.\ T.\ Nguyen, A.\ Yu.\ Grosberg, B.\ I.\ Shklovskii,
J. Chem. Phys. {\bf 113}, 1110 (2000).
\bibitem{Nguyen1} T.\ T.\ Nguyen, B.\ I.\ Shklovskii, 
Physica A {\bf 293}, 324 (2001).
\bibitem{Nguyen2} T.\ T.\ Nguyen, B.\ I.\ Shklovskii,
J. Chem. Phys. {\bf 114}, 5905 (2001); {\bf 11X}, XXXX (2001).
\bibitem{Nguyen3} T.\ T.\ Nguyen, B.\ I.\ Shklovskii, 
Phys. Rev. E {\bf 64}, 0414XX (2001). 
\bibitem{Rubinstein} A. V. Dobrynin, A. Deshkovski and M. Rubinstein,
Macromolecules {\bf 34}, 3421 (2001).
\bibitem{Shklovskii} V.\ I.\ Perel and B.\ I.\ Shklovskii, Physica A 274,
446 (1999); B.\ I.\ Shklovskii, Phys. Rev. E {\bf 60}, 5802
(1999). 
\bibitem{RMP} T. T. Nguyen, A. Yu. Grosberg, and B. I. Shklovskii,
cond-mat/0105140 (submitted to Rev. of Mod. Phys.).
\bibitem{Tao} R. Tao, D.\ J.\ Thouless, Phys. Rev. B {\bf 28}, 1142
(1983). 
\bibitem{Braz} S. Brazovskii, N. Kirova, JETP Lett. {\bf 33}, 4, (1981) 
\bibitem{KabanovD} V. A. Kabanov, V. G. Sergeyev, O. A. Pyshkina,
A. A. Zinchenko, A. B. Zezin, J. G. H. Joosten, J. Brackman,
and K. Yoshikawa, Macromolecules {\bf 33}, 9587 (2000); 
H.\ M.\ Evans, A.\ Ahmad, T. Pfohl,
A.\ Martin, C.\ R. Safinya Bull. APS, {\bf 46} 391 (2001).
\end{references}
\end{document}